\documentclass[twocolumn,prd,nofootinbib]{revtex4}
\usepackage{amsmath}
\usepackage{graphicx}

\begin{document}

\title{Geometrodynamics in a spherically symmetric, static crossflow of null
dust}
\author{Zsolt Horv\'{a}th, Zolt\'{a}n Kov\'{a}cs, L\'{a}szl\'{o} \'{A}.
Gergely}
\affiliation{Departments of Theoretical and Experimental Physics, University of Szeged,
Szeged 6720, D\'{o}m t\'{e}r 9, Hungary}

\begin{abstract}
The spherically symmetric, static spacetime generated by a crossflow of
non-interacting radiation streams, treated in the geometrical optics limit
(null dust) is equivalent to an anisotropic fluid forming a radiation
atmosphere of a star. This reference fluid provides a preferred / internal
time, which is employed as a canonical coordinate. Among the advantages we
encounter a new Hamiltonian constraint, which becomes linear in the momentum
conjugate to the internal time (therefore yielding a functional Schr\"{o}%
dinger equation after quantization), and a strongly commuting algebra of the
new constraints.
\end{abstract}

\maketitle

\section{Introduction}

Beside the covariant description of general relativity, a Hamiltonian
formalism of gravity based on the existence of foliations for globally
hyperbolic space-times was developed by Arnowitt, Deser and$\,$Misner
(hereafter ADM) \cite{ArnowittDeserMisner}. In this approach the canonical
coordinates are the components of the induced metric on the 3-leaves of the
foliation, while the canonical momenta are related in a simple way to the
extrinsic curvature of these spatial hypersurfaces. Gravitational evolution
is therefore quoted as geometrodynamics. The freedom to perform coordinate
transformations on the leaves of the foliation leads to the diffeomorphism
(or momentum) constraints. The true dynamics is encompassed in the so-called
Hamiltonian constraint. These four constraints (per point) form a Dirac
algebra \cite{Dirac}, which is not a true algebra in the mathematical sense,
as its closure is obstructed by the appearance of the induced metric in the
Poisson brackets of the Hamiltonian constraints.

The problem of time in canonical gravity was reviewed by Isham in Ref. \cite%
{Isham}. Approaches for introducing the concept of time are of three types:
time is either identified before or after quantization, or in certain
approaches time plays no fundamental role at all. In what follows, we are
interested in identifying time at the classical level. Time is not
preselected by any Hamiltonian description of gravity, there are infinitely
many ways to choose the time (many-fingered time formalism) \cite%
{IshamKuchar}. Despite this ambiguity, in certain cases it is possible to
select a preferred time function, by either imposing coordinate conditions 
\cite{KucharTorre} or by filling space-time with an adequate reference fluid 
\cite{BrownKuchar}, \cite{Brown} and letting gravity to evolve in the
(proper) time of the chosen reference fluid.

In certain cases new canonical variables can be introduced, providing new
constraints for gravity \cite{Markopoulou}, \cite{KucharRomano}, \cite%
{Kouletsis}. Among the advantages we count that the Dirac algebra transforms
into a true algebra and the quantization of the Hamiltonian constraint,
usually leading to the Wheeler-de Witt equation (which has no linear space
of solutions), rather gives a Schr\"{o}dinger equation. This program has
been particularly successful for incoherent dust, as presented by Brown and
Kucha\v{r} in Ref. \cite{BrownKuchar}.

A similar formalism \cite{BicakKuchar} was by Bi\v{c}\'{a}k and Kucha\v{r}
applied for null dust, the geometrical optics approximation to
non-gravitational radiation. Null dust however provides no natural time
function, basically because, unlike the congruence of the incoherent dust
particles, null world lines have no natural parametrization. While for
ordinary dust the Hamiltonian- and supermomentum constraints depend on four
pairs of canonical variables associated with the proper time and the
comoving coordinate frame of the dust, the constraint equations for null
dust contain only three pairs of comoving coordinates.

The quantum theory of gravitational collapse can be modelled in the most
simple spherically symmetric case by a collapsing thin shell of null dust 
\cite{HajicekKiefer},\ \cite{Hajicek}. A second null dust shell can be
introduced in the model in order to test the quantum behaviour of the
geometry\ induced by the first shell. Motivated by certain problems in the
above scenario, the canonical formalism in the presence of a null dust has
been recently extended to the case of two cross-flowing, non-interacting
null dust streams in a spherically symmetric space-time by Bi\v{c}\'{a}k and
H\'{a}j\'{\i}\v{c}ek \cite{BicakHajicek}. This formalism combines
ingredients of the canonical formalisms developed for null dust \cite%
{BicakKuchar} with elements of the geometrodynamics of the Schwarzschild
space-time \cite{Kuchar}, developed by Kucha\v{r}. The lack of \ a
time-standard for a single null dust however deprived the canonical
formalism of the cross-streaming null dust from a time-standard as well.
This is because the starting point of the canonical description \cite%
{BicakHajicek} is simply the sum of the spherically reduced Einstein-Hilbert
action for gravity and \ two pieces of the null dust action, also reduced by
spherically symmetry. The null dust variables are therefore doubled, without
any of them becoming an internal time. The basic assumption of Ref. \cite%
{BicakHajicek} is that the cross-flowing null dust streams interact only
gravitationally, therefore the energy-momentum tensors of the components are
conserved separately. The analysis of the equations of motion provides two
pairs of integrals of motion (per point), one pair for each null dust
component. Unfortunately the Hamiltonian density could not be explicitly
expressed in terms of these quantities, except in the case when one of the
null dust components is switched off. \ In this case the action can be
transformed such that the matter part of the Liouville form contains the
integrals of motion associated to the null dust component in question.

The formalism derived in Ref. \cite{BicakHajicek} is valid for certain known
spherically symmetric space-times, for example the Vaidya space-time,
describing the one-component null dust \cite{Vaidya}, and the \textit{static}
space-time found in Ref. \cite{Gergely} by one of the present authors. The
latter space-time represents the geometry in the presence of a static
cross-flow of non-interacting null dust streams. Although it is
asymptotically non-flat and it has a central naked singularity, it can be
conveniently interpreted as the radiation atmosphere of a star. A second
interpretation presented in Ref. \cite{Gergely} is of a 2-dimensional
dilatonic model, in the presence of a pair of 2-dimensional scalar fields.
While the dilaton is the square of the radial coordinate, the scalar fields
are related to the energy densities of the null dust streams. The third
interpretation, based on previous work of Letelier \cite{Letelier}, is of an
anisotropic fluid, with radial pressure equal to its energy density and no
tangential pressures. The static solution \cite{Gergely} has a homogenous
counterpart \cite{Gergely1999}, which can be interpreted as a
Kantowski-Sachs type cosmology. These two space-times obey a unicity
theorem, as they are the only spherically symmetric solutions of the
Einstein equation in the presence of a cross-flow of null dust streams with
an additional (fourth) Killing vector \cite{Gergely1999}. Interestingly, for
null dust streams with negative energy density, wormhole space-times emerge 
\cite{Hayward}, \cite{Gergely2002}.

The anisotropic fluid interpretation of the static space-time the cross-flow
of null dust streams with positive energy densities is particularly
important for our purposes. The physical model of the anisotropic fluid 
\textit{has} a preferred time, which is the time elapsed in the rest frame
of the fluid. This suggests that in contrast with the single null dust
model, for the two component null dust an internal time formalism can be
constructed. In this paper we will explicitly construct the matter action
for the static configuration of non-interacting null dust streams in terms
of suitable variables, containing the internal time singled out uniquely by
the cross-flow of null dust.

In Sec. II we summarize the basic ingredients necessary for the purposes of
the present work. We present:

\begin{description}
\item[(A)] the canonical formalism of ordinary incoherent dust \cite%
{BrownKuchar}, with special emphasize on how the proper choice of the
internal time allows us to introduce a set of new constraints for gravity,
such that the new super-Hamiltonian constraint becomes \textit{linear} in
the canonical momentum conjugate to the internal time;

\item[(B)] the geometrodynamics of the spherically symmetric static vacuum 
\cite{Kuchar}, with special emphasize on the introduction of geometrically
motivated canonical variables (including the Schwarzschild mass) in the
gravitational sector;

\item[(C)] the spherically symmetric, static space-time with crossflowing
null dust streams \cite{Gergely} and

\item[(D)] the anisotropic fluid interpretation of the cross-flow of
non-interacting null dust streams \cite{Letelier}, which provides the
internal time for the two component null dust.
\end{description}

In Sec. III we introduce an action functional of three scalar fields
characterizing the static cross-flow of null dust minimally coupled to
gravity. We show that variation with respect to the metric together with the
equations of motion reproduces the energy-momentum tensor of two
non-interacting radiation streams. Two pairs of conservation equations for
the rest mass currents and the momentum currents also emerge.

In Sec. IV we derive the contribution of the two null dust streams to the
super-Hamiltonian and diffeomorphism constraints. Then we fulfill the
program of replacing the total super-Hamiltonian and diffeomorphism
constraints by an equivalent set, in which both momenta conjugate to the
temporal and radial canonical variables appear \textit{linearly}. We also
prove that the new constraints form an Abelian algebra.

Sec V. contains a discussion of the falloff conditions the gravitational
variables, the lapse and the shift should obey.

In Sec. VI we compare our findings with the results presented in Ref. \cite%
{BicakHajicek} and we show that similar techniques can be employed in the
more generic context of Ref. \cite{BicakHajicek} as well. We also underline
the connections between our canonical variables and those employed in Ref. 
\cite{BicakHajicek}, specified for the static case.

Finally in Sec. VII we summarize our results.

\section{Preliminaries}

In this section we present a more technical summary of the results of Refs. 
\cite{BrownKuchar}, \cite{Kuchar}, \cite{Gergely} and \cite{Letelier}\
needed later on in the paper.

\subsection{Geometrodynamics of space-times with ordinary dust}

The space-time action of ordinary dust was constructed by Brown and Kucha%
\v{r} \cite{BrownKuchar} from eight scalar fields $Z^{k},W_{k},T,M$ \ ($%
k=1,2,3$) \ minimally coupled to the space-time metric $^{(4)}g_{ab}$:%
\begin{align}
& S^{D}\left[ T,Z^{k},M,W_{k};^{(4)}g_{ab}\right]  \notag \\
& =-\frac{1}{2}\int d^{4}x\sqrt{-{}^{(4)}g}M(U_{a}U^{a}+1)~,  \label{dustact}
\end{align}
The four-velocity $U_{a}$ is expressed as the Pfaff form of seven scalar
fields. 
\begin{equation}
U_{a}=-T_{,a}+W_{k}Z^{k}{}_{,a}\;.  \label{UA}
\end{equation}
The equations of motion are%
\begin{align}
0 & =\frac{\delta S^{D}}{\delta M}=-\frac{1}{2}\sqrt{-{}^{(4)}g}%
(U_{a}U^{a}+1)\;,  \label{DM} \\
0 & =\frac{\delta S^{D}}{\delta W_{k}}=-\sqrt{-{}^{(4)}g}MZ^{k}{}_{,a}U^{a}%
\;,  \label{DW} \\
0 & =\frac{\delta S^{D}}{\delta T}=-(\sqrt{-{}^{(4)}g}MU^{a})_{,a}\;,
\label{DT} \\
0 & =\frac{\delta S^{D}}{\delta Z^{k}}=-(\sqrt{-{}^{(4)}g}%
MW_{k}U^{a})_{,a}\;.  \label{DZ}
\end{align}
According to Eq. (\ref{DW}) the three vector fields $Z^{k}$ are constant
along the flow lines of $U^{a}$ (they can be interpreted as comoving
coordinates for the dust.)\ \ Eq. (\ref{DM}) shows that the four-velocity $%
U^{a}$ is a unit time-like vector field. Eq. (\ref{DT}) allows us to
interpret $M$ as the rest mass density of the dust and it represents mass
conservation. \ Eq. (\ref{DZ}) can be interpreted as the momentum
conservation law. From Eqs. (\ref{UA}), (\ref{DM})\ and (\ref{DW}) \ it is
straightforward to deduce that $T$ is the proper time along the dust world
lines, measured between a fiducial hypersurface $T=0$ and an arbitrary
hypersurface with constant $T$. The dust energy-momentum tensor $T_{ab\text{ 
}}$can be found from the variation of the action (\ref{dustact}) with
respect to $^{(4)}g_{ab}$. From the conservation of \ $T_{ab\text{ \ }}$and $%
\ M$ it follows that the dust particles evolve along geodesics.

The Legendre transformed action is%
\begin{align}
& S^{D}\left[ T,Z^{k},P,P_{k},g_{ab},N,N^{a}\right]  \notag \\
& =\int dt{\displaystyle \int}d^{3}x(P\dot{T}+P_{k}\dot{Z}%
^{k}-NH_{{}}^{D}-N^{a}H_{a}^{D})\;\;,
\end{align}
where $g_{ab}$ denotes the induced metric on the leaves, $N$ and $N^{a}$ are
the lapse function and shift vectors, respectively, and the momenta $P$ and $%
P_{k}$ are conjugate to $T$ and $Z^{k}$. (The original variables $W_{k}$
were expressed in terms of $P$ and $P_{k}$.) The constraints are%
\begin{align}
H_{\bot}^{D} & =\frac{1}{2}\frac{P^{2}}{Mg^{1/2}}+\frac{1}{2}\frac{Mg^{1/2}}{%
P^{2}}\left( P^{2}+g^{ab}H_{a}^{D}H_{b}^{D}\right) \;\;, \\
H_{a}^{D} & =PT_{,a}+P_{k}Z_{,a}^{k}\;\;.  \label{dustimp}
\end{align}
The dependence of the Hamiltonian constraint on the variable $M$ \ is
spurious. This can be shown as follows. By varying the action with respect
to $M$ we obtain an algebraic expression from which $M$ \ can be given in
terms of the other variables. Substituting this into\ the Hamiltonian
constraint gives 
\begin{equation}
H_{\bot}^{D}=\sqrt{P^{2}+g^{ab}H_{a}^{D}H_{b}^{D}}\,\,,  \label{dustham}
\end{equation}
so the mass multiplier $M$ is eliminated from the action.

By employing that the \textit{total} (gravitational + dust) constraints have
to vanish, e.g.\ $H_{\bot}^{D}=-H_{\bot}^{G}$ and $H_{k}^{D}=-H_{k}^{G}$ on
the constraint hypersurface, and solving the constraints (\ref{dustham}), (%
\ref{dustimp}) with respect to the momenta, we can replace the old
constraints by an equivalent set. The new super-Hamiltonian constraint can
be cast into the form%
\begin{align}
H_{\uparrow} & =P+h[g_{ab,}p^{ab}]=0\;,  \label{dustlinear} \\
h & =-\sqrt{\left( H_{\bot}^{G}\right) {}^{2}-g^{ij}H_{i}^{G}H_{j}^{G}\;}~,
\label{gyokg}
\end{align}
where $p^{ab}$ are the momenta conjugate to $g_{ab}$. Similarly the new
supermomentum constraint is:%
\begin{align}
H_{\uparrow k} & =P_{k}+h_{k}[T,Z^{k},g_{ab,}p^{ab}]=0\;,
\label{dustlinear2} \\
h_{k} & =Z_{k}^{a}H_{a}^{G}-hT_{,a}Z_{k}^{a}~.
\end{align}
The quantization of the linearized constraint (\ref{dustlinear}) gives a Schr%
\"{o}dinger equation \cite{BrownKuchar}.

\subsection{Geometrodynamics of spherically symmetric static vacuum}

After the preliminary studies on the canonical formalism of the spherically
symmetric space-times \cite{BCMN}, a comprehensive analysis of Hamiltonian
dynamics for Schwarzschild black holes was given by Kucha\v{r} \cite{Kuchar}%
. In this section we summarize those results of \ his work which are
relevant for our purposes.\ 

The space-time was foliated by spherically symmetric leaves $\Sigma_{t}$
which were labelled by the parameter time $t$. The induced metric on these
3-leaves can be characterized by two metric functions $\Lambda$ and $R$, 
\begin{equation}
g_{ab}dx^{a}dx^{b}=\Lambda^{2}(t,r)dr^{2}+R^{2}(t,r)d\Omega^{2}\;\;,
\label{3metric}
\end{equation}
where $r$ is a space-like coordinate and $d\Omega^{2}$ is the line element
on the unit sphere. Under coordinate transformations $R$ behaves as a scalar
and $\Lambda$ as a scalar density. In the ADM decomposition of the
spherically symmetric geometry, the shift vector has a non-vanishing
component only in the radial direction, denoted with $N^{r}$, which together
with the lapse function $N$ depend solely on the variables $t$ and $r$.

The metric functions $R$ and $\Lambda$ are chosen as canonical coordinates
and their momenta, as derived in \cite{Kuchar}, are 
\begin{align}
P_{\Lambda} & =-N^{-1}R(\dot{R}-R^{\prime}N^{r})\;\;,  \notag \\
P_{R} & =-N^{-1}\left[ \Lambda(\dot{R}-R^{\prime}N^{r})+R(\dot{\Lambda }%
-(\Lambda N^{r})^{\prime})\right] \;\;.
\end{align}
The vacuum action for the spherically symmetric geometry can be written as%
\begin{align}
& S_{\Sigma}[g_{ab},N,N^{a}]  \notag \\
& =\int dt\int_{\Sigma_{t}}d^{3}x(\dot{\Lambda}P_{\Lambda}+\dot{R}%
P_{R}-NH_{\bot}^{G}-N^{r}H_{r}^{G})\;\;,  \label{SSig}
\end{align}
with super-Hamiltonian and supermomentum constraints%
\begin{gather}
H_{\bot}^{G}[R,\Lambda,P_{\Lambda},P_{R}]=\frac{1}{R}P_{R}P_{\Lambda}+\frac {%
1}{2R^{2}}\Lambda P_{\Lambda}^{2}  \notag \\
+\frac{1}{\Lambda}RR^{\prime\prime}-\frac{R}{\Lambda^{2}}R^{\prime}\Lambda^{%
\prime}+\frac{1}{2\Lambda}R^{\prime2}-\frac{1}{2}\Lambda \;\;,
\label{HGbotori} \\
H_{r}^{G}[R,\Lambda,P_{\Lambda},P_{R}]=P_{R}R^{\prime}-P_{\Lambda}^{\prime
}\Lambda\;\;.  \label{HGrori}
\end{gather}

There exists a canonical transformation, through which the only dynamical
characteristics of the Schwarzschild space-time, the Schwarzschild mass $M$
turns into a canonical variable. The new set of variables is $(M,\mathrm{R}%
;P_{M},P_{\mathrm{R}})$, where $M(t,r)$ is expressed in terms of the old
variables $(\Lambda,R;P_{\Lambda},P_{R})$ through the formula of the
Schwarzschild mass derived by Kucha\v{r}:%
\begin{equation}
M=\frac{1}{2}R^{-1}P_{\Lambda}^{2}-\frac{1}{2}\Lambda^{-2}RR^{\prime}+\frac {%
1}{2}R\;\;.
\end{equation}
The remaining part of the canonical transformation is:%
\begin{align}
P_{M} & =\Lambda P_{\Lambda}\left( 1-\frac{2M}{R}\right) ^{-1}R^{-1}\;\;, 
\notag \\
\mathrm{R} & =R\;\;,  \notag \\
P_{\mathrm{R}} & =P_{R}-\frac{1}{2}R^{-1}\Lambda P_{\Lambda}-\frac{1}{2}%
\left( 1-\frac{2M}{R}\right) ^{-1}R^{-1}\Lambda P_{\Lambda}  \notag \\
& -R^{-1}\Lambda^{-1}\left( 1-\frac{2M}{R}\right) ^{-1}  \notag \\
& \times\left[ \left( \Lambda P_{\Lambda}\right) ^{\prime}RR^{\prime
}-\left( \Lambda P_{\Lambda}\right) \left( RR^{\prime}\right) ^{\prime }%
\right] \;\;.
\end{align}
\ The second advantage of the new set of canonical variables is that the
momentum $P_{M}$ is the gradient $T^{\prime}$ of the Schwarzschild time (cf.
Eq. (80) in Ref. \cite{Kuchar}). The gravitational constraints (\ref%
{HGbotori}) and (\ref{HGrori}), written in terms of the new canonical
variables, become%
\begin{align}
H_{\bot}^{G}[M,\mathrm{R},P_{M},P_{\mathrm{R}}] & =-\left( 1-\frac {2M}{%
\mathrm{R}}\right) ^{-1}\frac{M^{\prime}\mathrm{R}^{\prime}}{\Lambda }+ 
\notag \\
& +\left( 1-\frac{2M}{\mathrm{R}}\right) \frac{P_{M}P_{\mathrm{R}}}{\Lambda}%
\;\;,  \label{HGbot} \\
H_{r}^{G}[M,\mathrm{R},P_{M},P_{\mathrm{R}}] & =P_{\mathrm{R}}\mathrm{R}%
^{\prime}+P_{M}^{\prime}M\;\;,  \label{HGr}
\end{align}
where $\Lambda$ rather than being a canonical variable, is only a shorthand
notation for the following expression of the new canonical variables%
\begin{equation}
\Lambda=\left( 1-\frac{2M}{\mathrm{R}}\right) ^{-1}M^{\prime}{}^{2}-\left( 1-%
\frac{2M}{\mathrm{R}}\right) P_{M}^{2}\;\;.  \label{Lambda}
\end{equation}

We will also introduce the canonical variable $M$ in the description of the
gravitational sector of the cross-streaming null-dust space-time.

We mention here the related result of Varadarajan \cite{Varadarajan}, who
has derived a transformation from the usually employed canonical variables
(induced metric + extrinsic curvature), to a set of new canonical variables,
which have the interpretation of Kruskal coordinates. This transformation is
regular on the whole space-time, including the horizon. The constraints
simplifies in such an extent, that those are equivalent to the vanishing of
the canonical momenta.

\subsection{The spherically symmetric, static space-time with crossflowing
null dust streams}

The static superposition of two non-interacting null dust streams
propagating along the null congruence $u^{a}$ and $v^{a}$ is characterized
by the energy-momentum tensor%
\begin{equation}
T_{ab}^{2ND}=\rho(u_{a}u_{b}+v_{a}v_{b})~,  \label{enimp}
\end{equation}
with 
\begin{equation}
u_{a}u^{a}=v_{a}v^{a}=0\;,\qquad u_{a}v^{a}\neq0\;.  \label{uuvvuv}
\end{equation}
The same time-independent energy density $\rho$ was chosen for both null
dust components in order to assure no net energy flow (static configuration).

The spherically symmetric, static space-time containing such a cross-flow of
two non-interacting null dust streams has been presented in Ref. \cite%
{Gergely}: 
\begin{equation}
ds^{2}=-2a\frac{e^{L^{2}}}{R\left( L\right) }\left[ dZ^{2}-R^{2}(L)dL^{2}%
\right] +R^{2}(L)d\Omega^{2}\;,  \label{ds2}
\end{equation}
where $Z$ and $L$ are the time and radial coordinates adapted to the
symmetry and $R$ is the following expression of the radial coordinate: 
\begin{align}
-R(L) & =a(e^{L^{2}}-2L\Phi_{B})\;\;,  \notag \\
\Phi_{B} & =B+\int^{L}e^{x^{2}}dx\;.  \label{RL}
\end{align}
Here $a$ is a positive constant and $B$ is a parameter.

The four-velocity null vectors of the null dust streams are then 
\begin{align}
u_{a} & =WZ_{,a}+RWL_{,a}\;,  \notag \\
v_{a} & =WZ_{,a}-RWL_{,a}\,,  \label{uv}
\end{align}
with 
\begin{equation}
W=\sqrt{\frac{ae^{L^{2}}}{R}}\;.  \label{W}
\end{equation}
The energy density becomes 
\begin{equation}
\rho=\left( 8\pi R^{2}W^{2}\right) ^{-1}\;\,.  \label{rho}
\end{equation}
The superposition of the in- and outgoing null dust streams can be
interpreted as an anisotropic fluid. This indicates that there may be a
possibility to use the same procedure as in the case of the incoherent dust
to obtain an internal time for the canonical dynamics of cross-flowing (but
otherwise non-interacting) null dust streams, minimally coupled to gravity.

\subsection{The anisotropic fluid interpretation of the cross-flow of
non-interacting null dust streams}

Letelier has shown that the energy-momentum tensor of two null dust streams
is equivalent with the energy-momentum tensor of a specific anisotropic
fluid \cite{Letelier}. As consequence, the source of the static, spherically
symmetric space-times (\ref{ds2}) can be interpreted as an anisotropic fluid
with radial pressure equaling its energy density and no tangential pressures:%
\begin{equation}
T_{ab}=\rho(U_{a}U_{b}+\chi_{a}\chi_{b})~.  \label{fluid}
\end{equation}
Here $\chi^{\alpha}$ is the (normalized) radial direction and $U^{\alpha}$
is the unit~four-velocity of the fluid particles, obeying 
\begin{equation*}
-U_{a}U^{a}=\chi_{a}\chi^{a}=1\;,\qquad U_{a}\chi^{a}=0\;.
\end{equation*}
They are related to the null vectors by%
\begin{equation}
U^{\alpha}=\frac{1}{\sqrt{2}}(u^{a}+v^{a})\;,\qquad\chi^{\alpha}=\frac {1}{%
\sqrt{2}}(u^{a}-v^{a})\;.  \label{Uchi}
\end{equation}
By employing Eqs. (\ref{uv}), we can also express the vector fields $U^{a}$
and $\chi^{a}$ in the coordinate basis defined by $Z$ and $L$: 
\begin{equation}
U_{a}=\sqrt{2}WZ_{,a}\;,\qquad\chi_{a}=\sqrt{2}RWL_{,a}\;\;.
\label{Uchiform}
\end{equation}
In the anisotropic fluid picture $\rho$ represents both the energy density
and the pressure, while no tangential pressure components to the spheres of
constant $L$\ are present. The fluid is isotropic only about a single point,
the origin.

\section{Action principle for the static, spherically symmetric cross-flow
of two non-interacting radiation streams}

A generic spherically symmetric space-time, in coordinates $(T$, $R$, $%
\theta $, $\varphi)$ is characterized by two metric functions $h$ and $f$ as:%
\begin{equation}
ds^{2}=-h(T,R)dT^{2}+f^{-1}(T,R)dR^{2}+R^{2}d\Omega^{2}\,.  \label{sphe}
\end{equation}
Let us introduce two scalar fields $Z\left( T\right) $ and $L\left( R\right) 
$, and the following advanced-type and retarded-type combinations of the
1-forms $dZ$ and $dL$, which span the ($T$, $R$) sector: 
\begin{align}
u_{a} & =WZ_{,a}+RWL_{,a}\;,  \notag \\
v_{a} & =WZ_{,a}-RWL_{,a}\,,\;  \label{uv1}
\end{align}
with $W$ given by Eq. (\ref{W}). Thus in this co-basis the 1-forms $u_{a}$
and $v_{a}$ do not have time-dependent components. They are entirely
expressed in terms of the two scalar fields $Z\,\ $and $L$ (as the
coefficient functions $W\,\ $and $R$ can be given in terms of $L$). Note
that the expressions (\ref{uv1})\ are identical with (and in fact motivated
by) Eqs. (\ref{uv}), but this time the scalars $Z\,\ $and $L$ are \textit{not%
} related to any exact solution, and in consequence the 1-forms $u_{a}$ and $%
v_{a}$ are not necessarily null for the generic spherically symmetric metric
(\ref{sphe}). They do have instead the same length: 
\begin{align}
u^{a}u_{a} & =v^{a}v_{a}  \notag \\
& =-W^{2}\left[ h^{-1}\left( \frac{dZ}{dT}\right) ^{2}-fR^{2}\left( \frac{dL%
}{dR}\right) ^{2}\right] \;.  \label{equallength}
\end{align}
We also note that 
\begin{equation}
u^{a}v_{a}=-W^{2}\left[ h^{-1}\left( \frac{dZ}{dT}\right) ^{2}+fR^{2}\left( 
\frac{dL}{dR}\right) ^{2}\right] \;.  \label{uvcontr}
\end{equation}

Let us define a dynamical system by the action:%
\begin{equation}
S^{2ND}[{}^{(4)}g_{ab},\rho,Z,L]=-\frac{1}{2}\int d^{4}x\sqrt{-{}^{(4)}g}%
\rho(u_{a}u^{a}+v_{a}v^{a})\;,  \label{S2ND}
\end{equation}
where $\rho\left( L\right) $ is a third scalar field. We do not know at this
stage, what is the dynamical system described by the action (\ref{S2ND}).

Variation of the action with respect to the metric gives the energy-momentum
tensor:%
\begin{align}
T^{ab} & =\frac{-2}{\sqrt{-{}^{(4)}g}}\frac{\delta S^{2ND}}{\delta{}%
^{(4)}g_{ab}}  \notag \\
& =\frac{1}{2}{}^{(4)}g^{ab}\rho(u^{c}u_{c}+v^{c}v_{c})+%
\rho(u^{a}u^{b}+v^{a}v^{b})\;,  \label{dS2NDdgab}
\end{align}
while the variation with respect to the coordinates $Z$, $L$, and the
parameter $\rho$ give the Euler-Lagrange equations:%
\begin{align}
0=\frac{\delta S^{2ND}}{\delta Z} & =-2[\sqrt{-{}^{(4)}g}\rho
W(u^{a}+v^{a})]_{,a}\;,  \label{dS2NDdZ} \\
0=\frac{\delta S^{2ND}}{\delta L} & =2\sqrt{-{}^{(4)}g}\rho
L(u^{a}u_{a}+v^{a}v_{a})  \notag \\
& -2R[\sqrt{-{}^{(4)}g}\rho W(u^{a}-v^{a})]_{,a}\;,  \label{dS2NDdL} \\
0=\frac{\delta S^{2ND}}{\delta\rho} & =-\frac{1}{2}\sqrt{-{}^{(4)}g}%
(u_{a}u^{a}+v_{a}v^{a})\;.  \label{dS2NDdrho}
\end{align}
In Eq. (\ref{dS2NDdL}) we have employed the relation $dW/dL=W\left(
2RL-dR/dL\right) /2R.$

Eq. (\ref{dS2NDdrho}) together with Eq.\ (\ref{equallength}) implies that
both $u^{a}$ and $v^{a}$ are null vectors. Then the energy-momentum tensor (%
\ref{dS2NDdgab}) reduces to%
\begin{equation}
T^{ab}=\rho(u^{a}u^{b}+v^{a}v^{b})\;,  \label{Tab}
\end{equation}
characterizing a non-interacting cross-flow (in the null directions $u^{a}$
and $v^{a}$) of null dust streams with energy density $\rho$.

One can define rest mass currents as in \cite{BrownKuchar}%
\begin{equation}
\mathcal{J}^{a}:=\sqrt{-{}^{(4)}g}\rho u^{a}\;,\qquad\mathcal{K}^{a}:=\sqrt{%
-{}^{(4)}g}\rho v^{a}\,,  \label{mascur}
\end{equation}
and momentum currents as%
\begin{equation}
\mathcal{J}_{L}^{a}:=W\mathcal{J}^{a}\;,\qquad\mathcal{K}_{L}^{a}:=W\mathcal{%
K}^{a}\,.  \label{momcur}
\end{equation}
In term of these Eq. (\ref{dS2NDdZ}) is a continuity equation for the net
flow of radiation:%
\begin{equation}
\nabla_{a}(\mathcal{J}_{L}^{a}+\mathcal{K}_{L}^{a})=0\;.  \label{nabaJa1}
\end{equation}
As the vectors $u^{a}$ and $v^{a}$ are null, Eq. (\ref{dS2NDdL}) simplifies
to%
\begin{equation}
\nabla_{a}(\mathcal{J}_{L}^{a}-\mathcal{K}_{L}^{a})=0\;,  \label{nabaJa2}
\end{equation}
implying that both momentum currents are conserved individually: 
\begin{equation}
\nabla_{a}\mathcal{J}_{L}^{a}=0\;,\qquad\nabla_{a}\mathcal{K}_{L}^{a}=0\;,
\label{conservation}
\end{equation}
as expected for non-interacting radiation fields.

We have shown that the action defined by Eqs. (\ref{uv1}) and (\ref{S2ND})
describes a cross-flow of non-interacting null dust streams in a static
configuration with energy density $\rho$. \ As the vectors $u^{a}$ and $%
v^{a} $ are null, we can partially normalize them as $u^{a}v_{a}=-1$. Also,
from Eq. (\ref{equallength}) we get $dZ/dT=\left( fh\right) ^{1/2}RdL/dR$.
Then Eq. (\ref{uvcontr}) allows to express both metric functions as%
\begin{align}
f^{-1} & =2ae^{L^{2}}R\left( \frac{dL}{dR}\right) ^{2}\;, \\
h & =\frac{2ae^{L^{2}}}{R}\left( \frac{dZ}{dT}\right) ^{2}\;,
\end{align}
By inserting these into the generic spherically symmetric metric (\ref{sphe}%
), we obtain the metric form (\ref{ds2}), however without the additional
information (\ref{RL}) and (\ref{rho}). In order to recover these, we need
the Einstein equations, derived from the sum of the Einstein-Hilbert action
and the cross-flowing null dust action (\ref{S2ND}). These are identical to
those presented in Ref. \cite{Gergely}, thus lead to the solution summarized
in Section II.C.

At the end of this section we note that the equivalent action in the
anisotropic fluid picture is 
\begin{equation}
S^{F}[^{(4)}g^{ab},\rho,Z,L]=-\frac{1}{2}\int d^{4}x\sqrt{-{}^{(4)}g}%
\rho(U_{a}U^{a}+\chi_{a}\chi^{a})\;,  \label{LD0}
\end{equation}
with $U_{a}$ and $\chi_{a}$ given by Eq. (\ref{Uchiform}) . Due to the
equivalence of the two interpretations, all equations are the same,
irrespective of they being derived from the cross-streaming null dust action
(\ref{S2ND}) or from the anisotropic fluid action (\ref{LD0}).

\section{Canonical formalism}

In this section we present the calculations yielding linearized constraints
for the two-component null dust, similar to Eqs. (\ref{dustlinear}) and (\ref%
{dustlinear2}) derived for ordinary dust.

\subsection{3+1 decomposition of the two null dust Lagrangian}

The ADM decomposition of any spherically symmetric metric yields \cite%
{Kuchar}: 
\begin{equation}
ds^{2}=-(N^{2}-\Lambda^{2}N^{r}{}^{2})dt^{2}+2\Lambda^{2}N^{r}drdt+\Lambda
^{2}dr^{2}+R^{2}d\Omega^{2},  \label{ds22}
\end{equation}
where $\Lambda$ and $R$ are the metric functions from the induced
line-element (\ref{3metric}) and ($t$, $r)$ are generic coordinates
orthogonal to the ($\theta,\,\varphi)$ sector . The variables $\rho$, $Z$, $%
L $ characterizing the radiation cross-flow thus depend on both coordinates: 
$\rho=\rho(t,r),$ $Z=Z(t,r)$ and $L=L(t,r)$. From Eq. (\ref{ds22}) $\sqrt{%
^{(4)}g}=N\sqrt{g}$ \ \ .

The (3+1)-split form of the Lagrangian density taken from the action (\ref%
{S2ND}) is 
\begin{gather}
L^{2ND}=\frac{a\sqrt{g}\rho W^{2}}{N}\left( \dot{Z}^{2}\!-\!2N^{r}\dot {Z}%
Z^{\prime}\!-\!\frac{N^{2}-\Lambda^{2}N^{r}{}^{2}}{\Lambda^{2}}%
Z^{\prime}{}^{2}\right)  \notag \\
+\frac{a\sqrt{g}\rho R^{2}W^{2}}{N}\left( \dot{L}^{2}-2N^{r}\dot{L}L^{\prime
}-\frac{N^{2}-\Lambda^{2}N^{r}{}^{2}}{\Lambda^{2}}L^{\prime}{}^{2}\right) \,.
\label{LD}
\end{gather}
The canonical momenta conjugate to the radiation variables $Z$ and $L$
become 
\begin{align}
P_{Z}:= & \frac{\partial L^{2ND}}{\partial\dot{Z}}=\frac{2a\sqrt{g}\rho W^{2}%
}{N}(\dot{Z}-N^{r}Z^{\prime})\;,  \notag \\
P_{L}:= & \frac{\partial L^{2ND}}{\partial\dot{L}}=\frac{2a\sqrt{g}\rho
R^{2}W^{2}}{N}(\dot{L}-N^{r}L^{\prime})\;,  \label{momenta}
\end{align}
or inverted with respect to the velocities we obtain%
\begin{align}
\dot{Z} & =\frac{N}{2a\sqrt{g}\rho W^{2}}P_{Z}+N^{r}Z^{\prime},  \notag \\
~~\dot{L} & =\frac{N}{2a\sqrt{g}\rho R^{2}W^{2}}P_{L}+N^{r}L^{\prime}.
\label{velocity}
\end{align}
By inserting the velocities only in one factor of the velocity-squared terms
of (\ref{LD}) we obtain the Lagrangian in the \textquotedblright already
Hamiltonian\textquotedblright\ form%
\begin{equation}
L^{2ND}=\dot{Z}P_{Z}+\dot{L}P_{L}-NH_{\bot}^{2ND}-N^{r}H_{r}^{2ND}\;,
\label{LagHam}
\end{equation}
where the Hamiltonian and momentum constraints associated with the
cross-flow of null dust streams are found to be%
\begin{align}
H_{\bot}^{2ND} & =\frac{1}{2a\sqrt{g}\rho W^{2}}\left( P_{Z}^{2}+\frac{%
P_{L}^{2}}{R^{2}}\right)  \notag \\
& +\frac{2a\sqrt{g}\rho W^{2}}{\Lambda^{2}}\left(
Z^{\prime}{}^{2}+R^{2}L^{\prime}{}^{2}\right) \,,  \label{HDbot} \\
H_{r}^{2ND} & =Z^{\prime}P_{Z}+L^{\prime}P_{L}\;.  \label{HDr}
\end{align}
Remarkably, the momentum constraint has the same form as the dust constraint
(\ref{dustimp}).

\subsection{Introduction of new dust constraints}

If we vary the dust action (\ref{LD}) with respect to the comoving density $%
\rho$ of the dust, we obtain%
\begin{equation}
\frac{\delta S^{2ND}}{\delta\rho}=-N\frac{\partial H_{\bot}^{2ND}}{%
\partial\rho}=0\;,
\end{equation}
from which $\rho$ can be expressed as 
\begin{equation}
2a\sqrt{g}\rho W^{2}=\Lambda\sqrt{\frac{P_{Z}^{2}+P_{L}^{2}/R^{2}}{Z^{\prime
2}+R^{2}L^{\prime2}}}\;.
\end{equation}
By substituting this result into the Hamiltonian constraint (\ref{HDbot}),
we get 
\begin{align}
& H_{\bot}^{2ND}  \notag \\
& =\frac{2}{\Lambda}\sqrt{R^{2}(P_{Z}L^{\prime})^{2}+\frac{(P_{L}Z^{\prime
})^{2}}{R^{2}}+(P_{Z}Z^{\prime})^{2}+(P_{L}L^{\prime})^{2}}\,.  \label{ROOT}
\end{align}
Since (\ref{HDr}) implicates that the last two terms below the root appear
in $\left( H_{r}^{2ND}\right) ^{2}$, we eliminate them from (\ref{ROOT}).
The final form of the Hamiltonian constraint is%
\begin{align}
& H_{\bot}^{2ND}  \notag \\
& =2\sqrt{\left( \frac{P_{Z}L^{\prime}R}{\Lambda}-\frac{P_{L}Z^{\prime}}{%
\Lambda R}\right) ^{2}+g^{rr}H_{r}^{2ND}H_{r}^{2ND}}\;.  \label{HDbot2}
\end{align}
We note that in the spherically symmetric case the momentum constraint $%
H_{r}^{2ND}$ can also be brought to a square root form. From Eq. (\ref%
{3metric}) and (\ref{HDbot2}) we have%
\begin{equation}
H_{r}^{2ND}=\sqrt{-\left( P_{Z}L^{\prime}R-\frac{P_{L}Z^{\prime}}{R}\right)
^{2}+\frac{1}{4}(\Lambda H_{\bot}^{2ND})^{2}}\;.
\end{equation}
Eq. (\ref{HDbot2}) is of similar form to the Hamiltonian constraint of the
incoherent dust derived in \cite{BrownKuchar}.\ There is one difference,
namely that the Hamiltonian constraint (\ref{dustlinear}) of the incoherent
dust depends on the momenta conjugate to the 3-dimensional coordinate frame
variables only through the momentum constraint, while in (\ref{HDbot2}) $%
P_{L}$ appears both explicitly and through $H_{r}^{2ND}$. \ In spite of
this, we can still follow the algorithm of \cite{BrownKuchar}, as will
become transparent in the following.

The ADM decomposition of the total action leads to the super Hamiltonian and
super momentum constraints 
\begin{align}
H_{\bot}:= & H_{\bot}^{G}+H_{\bot}^{2ND}=0\;,  \label{Hbot} \\
H_{r}:= & H_{r}^{G}+H_{r}^{2ND}=0\;,  \label{Hr}
\end{align}
where the vacuum constraints $H_{\bot}^{G}$ and $H_{r}^{G}$ are expressed in
terms of the canonical variables $(M,\mathrm{R};P_{M},P_{\mathrm{R}})$ in
Eqs. (\ref{HGbot}) and (\ref{HGr}). By using Eqs. (\ref{HDr}), (\ref{Hbot})
and (\ref{Hr}) we can eliminate $P_{L}$ , $H_{\bot}^{2ND}$ and $H_{r}^{2ND}$
from Eq. (\ref{HDbot2}) to obtain%
\begin{align}
& -H_{\bot}^{G}  \notag \\
& =2\sqrt{\left( \frac{\left( Z^{\prime2}+L^{\prime2}R^{2}\right)
P_{Z}+Z^{\prime}H_{r}^{G}}{L^{\prime}\Lambda R}\right)
^{2}+g^{rr}H_{r}^{G}H_{r}^{G}}\;.  \label{MINHBOTG}
\end{align}
Then $P_{Z}$ can be separated from the other variables in Eq. (\ref{MINHBOTG}%
) :%
\begin{align}
H_{\uparrow Z} & =P_{Z}+h_{Z}[M,\mathrm{R},Z,L,P_{M},P_{\mathrm{R}}]=0\;\;, 
\notag \\
h_{Z} & =\frac{L^{\prime}\Lambda\mathrm{R}h+Z^{\prime}H_{r}^{G}}{Z^{\prime
2}+L^{\prime2}\mathrm{R}^{2}}~~.  \label{LINEARZ}
\end{align}
From (\ref{LINEARZ}) we know%
\begin{equation}
P_{Z}=-h_{Z}=-\frac{L^{\prime}\Lambda\mathrm{R}h+Z^{\prime}H_{r}^{G}}{%
Z^{\prime2}+L^{\prime2}\mathrm{R}^{2}}~.  \label{Pz}
\end{equation}
Here we used the notation (\ref{gyokg}).

By using (\ref{HDr}) and (\ref{Pz}), the constraint (\ref{Hr}) can be
written as%
\begin{align}
0 & =H_{r}:=H_{r}^{G}+H_{r}^{2ND}  \notag \\
& =H_{r}^{G}+Z^{\prime}P_{Z}+L^{\prime}P_{L}  \notag \\
& =H_{r}^{G}-Z^{\prime}\frac{L^{\prime}\Lambda Rh+Z^{\prime}H_{r}^{G}}{%
Z^{\prime2}+L^{\prime2}R^{2}}+L^{\prime}P_{L}~,  \label{startpoint}
\end{align}
Which gives%
\begin{equation}
0=P_{L}+\frac{-Z^{\prime}\Lambda Rh+H_{r}^{G}L^{\prime}R^{2}}{Z^{\prime
2}+L^{\prime2}R^{2}}~~.  \label{consteq}
\end{equation}
We will denote the constraint (\ref{consteq}) by $H_{\uparrow L}$.%
\begin{align}
H_{\uparrow L} & =P_{L}+\pi_{L}[M,\mathrm{R},Z,L,P_{M},P_{\mathrm{R}}]=0\;, 
\notag \\
\pi_{L} & =\frac{-Z^{\prime}\Lambda\mathrm{R}h+L^{\prime}\mathrm{R}%
^{2}H_{r}^{G}}{Z^{\prime2}+L^{\prime2}\mathrm{R}^{2}}\;\;.  \label{LINEARL}
\end{align}
Thus we have obtained a new, more convenient set of super-Hamiltonian
constraint $H_{\uparrow Z}$ and supermomentum constraint $H_{\uparrow L}$.
Both linearized constraints contain exactly one null dust momentum. The
Dirac algebra of the old constraints turns into an Abelian algebra of the
new constraints: 
\begin{equation}
\left\{ H_{\uparrow J}(r),H_{\uparrow J^{\prime}}(r^{\prime})\right\} =0\,\,,
\end{equation}
where $H_{\uparrow J}=\left( H_{\uparrow Z},H_{\uparrow L}\right) \,.$ This
feature is similar to the case of the one-component ordinary dust \cite%
{BrownKuchar}, and in fact the proof proceeds exactly in the same way.
Following \cite{BrownKuchar} first we note that the Poisson brackets of the
new constraints must vanish, at least weakly (on the constraint
hypersurface). However, due to the linearity of the constraints (\ref%
{LINEARZ}), (\ref{LINEARL}) in the momenta $P_{Z},\,P_{L}$, the brackets do
not depend on any of $P_{Z},\,P_{L}$. But then there is no way the
constraints (\ref{LINEARZ}), (\ref{LINEARL}) would help in turning into zero
the Poisson brackets. Therefore they have to strongly vanish.

\section{Falloff of the canonical variables}

\subsection{Falloff conditions for the eternal Schwarzschild black hole}

The proof of Kucha\v{r} in Ref. \cite{Kuchar} that the mapping ($\Lambda ,~R%
\mathrm{,}~P_{\Lambda },~P_{R})\rightarrow (M,~\mathrm{R,}~P_{M},~P_{\mathrm{%
R}})$ is a canonical transformation in the gravitational sector relies on
the check that the difference of the Liouvillle forms is an exact form. This
translates to show that the expression 
\begin{equation}
\mathcal{B}\left( r\right) =\frac{1}{2}R\delta R\ln \left\vert \frac{%
RR^{\prime }+\Lambda P_{\Lambda }}{RR^{\prime }-\Lambda P_{\Lambda }}%
\right\vert  \label{boundary}
\end{equation}%
vanishes on the boundaries of the domain of integration. For the eternal
Schwarzschild black hole discussed there, the desired behaviour was assured
at $r\rightarrow \pm \infty $ by imposing suitable falloff conditions for
the canonical variables, based on the treatment of Beig and O'Murchadha \cite%
{BeigOMurchadha}. The proper falloff of the variables $\Lambda ,~R\mathrm{,}%
~P_{\Lambda },~P_{R}$, Killing time $T$, lapse function $N$ and shift $N^{r}$%
, given by Eqs. (49)-(55) of Ref. \cite{Kuchar} assure that the Kucha\v{r}
mapping is a canonical transformation.

\subsection{Falloff conditions for $r\rightarrow0$ in flat space-time}

H\'{a}j\'{\i}\v{c}ek and Kiefer have studied the evolution of a spherically
symmetric null dust shell in the space-time generated by an other
spherically symmetric null dust shell \cite{HajicekKiefer}. The (innermost)
region surrounded by the incoming null shell is Minkowski. In order to avoid
the occurrence of a conical singularity at $r=0$, following the method
developed for cylindrical gravitational waves \cite{cylindrical}, they have
imposed boundary conditions on both the coordinates and their spatial
derivatives at the regular centre. Based on these, Bi\v{c}\'{a}k and H\'{a}j%
\'{\i}\v{c}ek \cite{BicakHajicek} have shown that the boundary term (\ref%
{boundary}) also vanishes at $r\rightarrow 0$.

Louko, Whiting and Friedman have discussed the Hamiltonian dynamics of a
thin (distributional) null-dust shell under both sets of boundary
conditions: first at the two spatial infinities $r\rightarrow \pm \infty $
of the Kruskal-like manifold and second at $r\rightarrow 0$ and $%
r\rightarrow \infty $ \cite{LoukoWhitingFriedman}. In the latter case, the
falloff conditions at $r\rightarrow 0$ for the canonical variables, lapse
and shift in the flat geometry within the null shell are given by their
system of Eqs. (7.1):%
\begin{align}
\Lambda (t,r)& =\Lambda _{0}+\mathcal{O}(r^{2})\;,  \notag \\
R(t,r)& =R_{1}r+\mathcal{O}(r^{3})\;,  \notag \\
P_{\Lambda }(t,r)& =P_{\Lambda _{2}}r^{2}+\mathcal{O}(r^{4})\;,  \notag \\
P_{R}(t,r)& =P_{R_{1}}r+\mathcal{O}(r^{3})\;,  \notag \\
N\left( t,r\right) & =N_{0}+\mathcal{O}(r^{2})\;,  \notag \\
N^{r}\left( t,r\right) & =N_{1}^{r}r+\mathcal{O}(r^{3})\;,  \label{falloff0}
\end{align}%
where $\Lambda _{0},~R_{1},~P_{\Lambda _{2}},~P_{R_{1}},~N_{0}$ and $%
N_{1}^{r}$ are functions of time. With these falloffs, the expression $%
\mathcal{B}\left( 0\right) $ vanishes, in accordance with the conclusion of
Ref. \cite{BicakHajicek}.

Given the falloff behaviors (\ref{falloff0}), all terms in the vacuum
gravitational super-Hamiltonian constraint (\ref{HGbotori}) are $\mathcal{O}%
\left( r^{2}\right) $, with two exceptions: $R^{\prime ~2}/2\Lambda
=R_{1}^{2}/2\Lambda _{0}+\mathcal{O}\left( r^{2}\right) $ and $-\Lambda
/2=-\Lambda _{0}/2+\mathcal{O}\left( r^{2}\right) $. Therefore 
\begin{equation}
H_{\bot }^{G}=\frac{R_{1}^{2}-\Lambda _{0}^{2}}{2\Lambda _{0}}+\mathcal{O}%
\left( r^{2}\right) ~.  \label{ham0}
\end{equation}%
The leading term vanishes for 
\begin{equation}
R_{1}=\Lambda _{0}~.  \label{cond1}
\end{equation}%
For this choice, the \ falloff conditions obey the vacuum gravitational
super-Hamiltonian constraint$.$ The gravitational super-momentum constraint (%
\ref{HGrori}), in turn behaves as 
\begin{equation}
H_{r}^{G}=-2P_{\Lambda _{2}}\Lambda _{0}~r+\mathcal{O}\left( r^{2}\right) ~.
\label{mom0}
\end{equation}%
Thus, the falloff conditions are consistent with the vacuum constraints.
They are also preserved in time, as noted in \cite{LoukoWhitingFriedman},
but we will show that only for 
\begin{equation}
P_{\Lambda _{2}}=0~.  \label{cond2}
\end{equation}%
This can be seen from the following argument \ The time-evolution of the
super-Hamiltonian and super-momentum constraints are linear combinations of
the constraints and their covariant derivatives on the leaves:%
\begin{align}
\dot{H}_{\bot }^{G}& =2H_{r}^{G}D^{r}N-2NKH_{\bot }^{G}  \notag \\
& +N^{r}D_{r}H_{\bot }^{G}+ND_{r}H^{Gr}~,  \notag \\
\dot{H}_{r}^{G}& =2H_{\bot }^{G}D_{r}N-NKH_{r}^{G}+H_{r}^{G}D_{r}N^{r} 
\notag \\
& +ND_{r}H_{\bot }^{G}+N^{r}D_{r}H_{r}^{G}~.  \label{constrevol}
\end{align}%
Here $K=\Lambda ^{-1}R^{-2}\left( RP_{R}-\Lambda P_{\Lambda }\right)
+2R^{-2}P_{\Lambda }$ is the trace of the extrinsic curvature of the leaves $%
\Sigma _{t}$., given in Ref. \cite{Kuchar}. From the falloff conditions (\ref%
{falloff0}) we obtain%
\begin{equation}
K=\Lambda _{0}^{-1}R_{1}^{-2}\left( R_{1}P_{R_{1}}+\Lambda _{0}P_{\Lambda
_{2}}\right) +\mathcal{O}\left( r\right) ~,
\end{equation}%
Thus the terms proportional to the gravitational constraints, whether they
contain $K$ or not, will decay at least as $\mathcal{O}\left( r\right) $ and 
$\mathcal{O}\left( r^{2}\right) $, respectively (provided $R_{1}=\Lambda _{0}
$ was chosen). Problems could aarise only from the terms containing
derivatives of the constraints. The falloff conditions (\ref{falloff0}) and
the covariant derivatives of the scalar and vector densities $H_{\bot }^{G}$
and $H_{r}^{G}$ give at $r\rightarrow 0$%
\begin{align}
N^{r}D_{r}H_{\bot }^{G}& =-N_{1}^{r}\frac{R_{1}^{2}-\Lambda _{0}^{2}}{%
\Lambda _{0}^{2}}+\mathcal{O}\left( r\right) ~,  \notag \\
ND_{r}H^{Gr}& =-2N_{0}\frac{P_{\Lambda _{2}}}{\Lambda _{0}}+\mathcal{O}%
\left( r\right) ~,  \notag \\
ND_{r}H_{\bot }^{G}& =-N_{0}\frac{R_{1}^{2}-\Lambda _{0}^{2}}{2\Lambda
_{0}^{3}}\left( 2r^{-1}+\Lambda _{0}^{-1}\right) +\mathcal{O}\left( r\right)
~,  \notag \\
N^{r}D_{r}H_{r}^{G}& =-2N_{1}^{r}\Lambda _{0}P_{\Lambda _{2}}r+\mathcal{O}%
\left( r^{2}\right) ~,  \label{constrevolterms}
\end{align}%
Therefore%
\begin{align}
\dot{H}_{\bot }^{G}& =-N_{1}^{r}\frac{R_{1}^{2}-\Lambda _{0}^{2}}{\Lambda
_{0}^{2}}-2N_{0}\frac{P_{\Lambda _{2}}}{\Lambda _{0}}+\mathcal{O}\left(
r\right) ~,  \notag \\
\dot{H}_{r}^{G}& =-N_{0}\frac{R_{1}^{2}-\Lambda _{0}^{2}}{2\Lambda _{0}^{3}}%
\left( 2r^{-1}+\Lambda _{0}^{-1}\right) +\mathcal{O}\left( r\right) ~.
\label{consistency}
\end{align}%
By chosing the condition (\ref{cond1}), the expression for $\dot{H}_{r}^{G}$
will decay as $\mathcal{O}\left( r\right) $. As we exclude the possibility $%
N_{0}=0$ (which would froze time evolution at $r=0$), the only possibility
remaining for a proper decay of $\dot{H}_{\bot }^{G}$ is to set $P_{\Lambda
_{2}}=0$, which completes our proof.

\subsection{Falloff conditions for the radiative atmosphere of a star}

\begin{figure}[tbp]
\includegraphics[height=9cm]{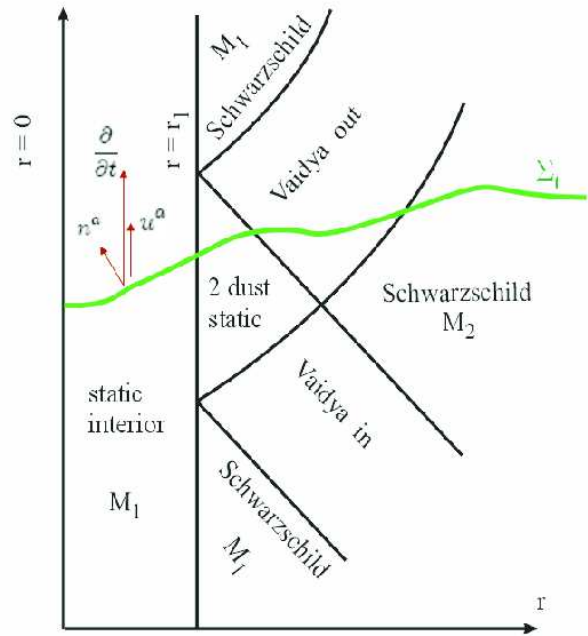}
\caption{(Color online) The static star, incoming and outgoing radiation
zones, crossflowing null dust region and three exterior Schwazschild
domains. The normal vector $n^{a}$ of the foliation leaf $\Sigma _{t}$ is
different from the fluid 4-velocity $u^{a}$. The foliation is chosen so that
time evolution in the static star region proceeds in the direction of the
fluid world-lines. }
\label{Fig1}
\end{figure}

Now we study the question, whether the Kucha\v{r} mapping ($\Lambda ,~R%
\mathrm{,}~P_{\Lambda },~P_{R})\rightarrow (M,~\mathrm{R,}~P_{M},~P_{\mathrm{%
R}})$ of the gravitational variables remains a canonical transformation in
the configuration discussed in this paper. In order to answer this question,
first we remark that the range of the cross-streaming null dust metric
parameter $B$ is restricted by $R>0$. This determines a lower boundary $%
L_{min}$ of $L$, corresponding to $R=0$. The quasi-local mass function $%
m\left( L\right) $ (for a definition see \cite{Gergely}) vanishes at some $%
L_{m=0}(a,B)$ and takes negative values below, in the interval $%
L_{min}<L<L_{m=0}$. Besides, for $L\rightarrow \infty $ the solution (\ref%
{ds2}) is not asymptotically flat. One can escape these unpleasant features
by cutting off the space-time between certain $L_{1}>L_{m=0}$ and an
appropriate high value $L_{2}>L_{1}$ and matching with appropriate metrics
across these boundaries (see Fig. \ref{Fig1}). The cross-streaming null dust
region (\ref{ds2}) is matched then from the interior with the interior
Schwarzschild solution, representing a static star with mass $M_{1}$,
whereas from the exterior it is bounded by incoming and outgoing Vaidya
regions, and it touches three exterior Schwarzschild regions in three points
(these are 2-spheres, if we take into account the angles $\theta ,~\varphi $%
). Therefore the solution (\ref{ds2}) is interpreted as a thick shell of
2-component radiation, created from the intersection of incoming and
outgoing thick radiation shells.

The intersection of the last incoming ray with the first outgoing ray is the
point (2-sphere) where the junction to the outermost Schwarzschild region
(characterized by mass $M_{2}$) is done. This region extends towards the
spatial infinity $i^{0}$. As the fluid region is bounded, only the proper
falloff at $i^{0}$ of the gravitational variables $M,~\mathrm{R,}~P_{M},~P_{%
\mathrm{R}}$ has to hold, as summarized in the first subsection of this
Section.

The situation is not so trivial on the other boundary, at $r\rightarrow 0$.
There, in contrast with the previous treatments of Refs. \cite{HajicekKiefer}%
, \cite{BicakHajicek} and \cite{LoukoWhitingFriedman}, we do not have
vacuum, but rather the center of a static star represented by the interior
Schwarzschild solution, where the falloff conditions are not yet known. The
line-element representing the gravitational field in the interior
Schwarzschild solution, 
\begin{align}
ds^{2}& =-\left( a-bF^{1/2}\right) ^{2}dt^{2}+F^{-1}dr^{2}+r^{2}d\Omega
^{2}~, \\
F\left( r\right) & =1-\frac{\kappa ^{2}\rho r^{2}}{3}~
\end{align}%
is generated by a perfect fluid with energy-momentum tensor%
\begin{equation}
T_{ab}=\left( \rho +p\right) u_{a}u_{b}+p~^{(4)}g_{ab}~,
\end{equation}%
where the energy density $\rho $ and pressure $p$ (\textit{with respect to
the 4-velocity }$u^{a}$\textit{\ of the fluid particles}.) are given as%
\begin{align}
\rho & =\text{const~,}  \notag \\
p& =\rho \frac{bF^{1/2}-a/3}{a-bF^{1/2}}~.
\end{align}%
Here $\kappa ^{2}=8\pi G$ and $a,~b$ are constants, chosen such that $p\geq
0 $.

As the canonical treatment of the interior Schwarzschild solution has not
been developed yet (and it is beyond the scope of the present paper), we
will impose the simplifying condition that the worldlines of the fluid
particles of the stellar material are along the time evolution vector $%
\partial /\partial t$. 
\begin{equation*}
\alpha u^{a}=\left( \frac{\partial }{\partial t}\right)
^{a}=Nn^{a}+N^{r}\Lambda ^{-1}\left( \frac{\partial }{\partial r}\right)
^{a}~,
\end{equation*}%
where $\alpha \left( t,r\right) >0$ is a scaling function. From the
condition of normalization of the 4-velocity $u^{a}u_{a}=-1~$\ we obtain$%
~\alpha ^{2}=N^{2}-N^{r~2}$. This choice of the allowable foliations is in
accordance with the generic expectation, that whenever a reference fluid is
present in the system, it is advantageaus to introduce the parameter
associated with the world-lines of the reference fluid as time variable.
Outside the interior Schwarzschild region, the leaves $\Sigma _{t}$ are
still allowed to be arbitrary space-like hypersurfaces.

The energy density and energy current density of the fluid \textit{with
respect to the chosen foliation} become\textit{\ }%
\begin{align}
\mu & =T_{ab}n^{a}n^{b}=\left( \frac{N}{\alpha}\right) ^{2}\rho+\left( \frac{%
N^{r}}{\alpha}\right) ^{2}p~,  \notag \\
j_{r} & =T_{ab}n^{a}g_{r}^{b}=-\frac{NN^{r}}{\alpha^{2}}\Lambda\left(
\rho+p\right) ~.
\end{align}

With the falloff conditions (\ref{falloff0}) at $r\rightarrow0$ the
condition $\mathcal{B}$ $\rightarrow0$ will continue to hold, thus the Kucha%
\v{r} transformation is canonical. But are these falloff conditions
consistent with the constraints? In order to responde affirmatively, first
we note that for the fluid variables we have the following falloff conditions%
\begin{align}
p & =\frac{3b-a}{3\left( a-b\right) }\rho_{S}+\mathcal{O}\left( r^{2}\right)
~,  \notag \\
\alpha & =N_{0}+\mathcal{O}(r^{2})~,  \notag \\
\mu & =\rho_{S}+\mathcal{O}(r^{2})~,  \notag \\
j_{r} & =-\frac{2a\rho_{S}}{3\left( a-b\right) }N_{0}^{-1}N_{1}^{r}%
\Lambda_{0}r+\mathcal{O}(r^{2})~.  \label{falloff2}
\end{align}
These, together with $\sqrt{g}=\Lambda R^{2}\sin\theta$ imply that 
\begin{align}
H_{\bot}^{star} & =2\kappa^{2}\sqrt{g}\mu=\mathcal{O}\left( r^{2}\right) ~,
\label{ham1} \\
H_{r}^{star} & =2\kappa^{2}\sqrt{g}j_{r}=\mathcal{O}\left( r^{3}\right) ~,
\label{mom1}
\end{align}
which shows that the total constraints of gravity and fluid are obeyed for
the chosen falloffs on the boundary, provided the condition (\ref{cond1})
holds.

The last question to address is whether time evolution conserves these
falloffs. In order to see this we note that both $\dot{H}_{\bot}^{star}$ and 
$\dot{H}_{r}^{star}$ vanish for the interior Schwarzschild solution and for
the chosen class of foliations as $\mathcal{O}(r^{2})$ and $\mathcal{O}%
(r^{3})$, respectively. Thus, the falloff of the matter part of the
constraints is faster than the falloff of the gravitational part, given by
Eqs. (\ref{consistency}). Fulfilling the conditions (\ref{cond1}) and (\ref%
{cond2}) is sufficient for the consistency with the constraints in the
interior Schwarzschild solution..

Alternatively, if we do not insist on the interpretation of the
cross-streaming null dust space-time region as a radiation atmosphere of a
star, we can let the outgoing radiation to emerge from the origin and the
incoming component to be absorbed by the boundary at $r=0$. In this case
Cauchy surfaces can be chosen in such a way, that their boundary at $%
r\rightarrow0$ is in a flat space-time, as in Fig. 3. of Ref. \cite%
{BicakHajicek}. In this setup, the expression $\mathcal{B}$ again vanishes,
and the Kuchar transformation is proved to be canonical.

\section{Comparison with previous results}

In this Section we will establish the connections between the sets of
canonical variables employed in this paper and in Ref. \cite{BicakHajicek}.
In order to do this first we illustrate in Subsec. V.A that an internal time
can be introduced for a generic spherically symmetric crossflow of radiation
streams. We start from the variables employed in Ref. \cite{BicakHajicek}.
In Subsec. V.B we show that the connection of those variables with our
variables can be written up explicitly.

\subsection{Constraints of null dust crossflow}

Bi\v{c}\'{a}k \ and H\'{a}j\'{\i}\v{c}ek \cite{BicakHajicek} generalized the
canonical formulation of the one-component null dust, presented in Ref. \cite%
{BicakKuchar} for a two-component null dust, with the specification of
spherically symmetry. The gravitational part of their action was given by (%
\ref{SSig}), whereas the energy-momentum tensor has been chosen as 
\begin{equation}
T^{ab}=\frac{1}{4\pi}\left( l_{+}^{a}l_{+}^{b}+l_{-}^{a}l_{-}^{b}\right) \;,
\label{BHTENZOR}
\end{equation}
with 
\begin{equation}
l_{\pm}^{a}=\frac{\sqrt{\left| \Pi_{\pm}\Phi_{\pm}^{\prime}\right| }}{%
\Lambda R}\left[ n^{a}\pm\Lambda^{-1}\left( \frac{\partial}{\partial r}%
\right) ^{a}\right] \;
\end{equation}
being the four-velocity null vectors of the ingoing and outgoing null dust
streams. The latter were characterized by the canonical coordinates $%
\Phi_{+} $ , $\Phi_{-}$ and their conjugate momenta $\Pi_{+}$ , $\Pi_{-}$.
The unit normal to the leaves was denoted $n^{a}$. The canonical action of
the system became 
\begin{align}
& S^{T}[N,N^{r},\Lambda,R,\Phi_{+},\Phi_{-},P_{\Lambda},P_{R},\Pi_{+},\Pi
_{-}]  \notag \\
& =\int dt\int dr(P_{\Lambda}\dot{\Lambda}+P_{R}\dot{R}+\Pi_{+}\dot{\Phi}%
_{+}+\Pi_{-}\dot{\Phi}_{-}-  \notag \\
& -NH_{\bot}^{T}-N^{r}H_{r}^{T})\;\;,
\end{align}
with the super-Hamiltonian constraint 
\begin{align}
H_{\bot}^{T} & :=H_{\bot}^{G}+H_{\bot}^{BH}=0\;,  \notag \\
H_{\bot}^{BH} & =\frac{\left| \Pi_{+}\Phi_{+}^{\prime}\right| +\left|
\Pi_{-}\Phi_{-}^{\prime}\right| }{\Lambda},  \label{HBTOTBH}
\end{align}
and super-momentum constraint%
\begin{align}
H_{r}^{T} & :=H_{r}^{G}+H_{r}^{BH}=0\;,  \notag \\
H_{r}^{BH} & =\Pi_{+}\Phi_{+}^{\prime}+\Pi_{-}\Phi_{-}^{\prime}\,.
\label{HRTOTBH}
\end{align}
Following the convention of Ref. \cite{BicakHajicek}, we assume that $\Pi
_{+}\Phi_{+}^{\prime}$ $<$ 0 $<$ $\Pi_{-}\Phi_{-}^{\prime}$ . Thus we will
use 
\begin{equation}
H_{\bot}^{BH}=\frac{-\Pi_{+}\Phi_{+}^{\prime}+\Pi_{-}\Phi_{-}^{\prime}}{%
\Lambda}\;\;.
\end{equation}

The constraints (\ref{HBTOTBH}) and (\ref{HRTOTBH}) can be conveniently
combined as follows%
\begin{equation}
0=\frac{\Lambda H_{\bot}^{T}-H_{r}^{T}}{-2\Phi_{+}^{\prime}}=\Pi_{+}+\frac{%
-\Lambda H_{\bot}^{G}+H_{r}^{G}}{2\Phi_{+}^{\prime}}\;\;.  \label{LINEARPLUS}
\end{equation}
Similarly%
\begin{equation}
0=\frac{\Lambda H_{\bot}^{T}+H_{r}^{T}}{2\Phi_{-}^{\prime}}=\Pi_{-}+\frac{%
\Lambda H_{\bot}^{G}+H_{r}^{G}}{2\Phi_{-}^{\prime}}\;\;.  \label{LINEARMINUS}
\end{equation}
The new constraints (\ref{LINEARPLUS}) and (\ref{LINEARMINUS}) are analogous
to the previously introduced constraints (\ref{LINEARZ}) and (\ref{LINEARL})
in containing the canonical momenta only linearly.

There is nothing to prevent us in introducing square-root type new
constraints by properly transforming the constraints (\ref{LINEARPLUS}), (%
\ref{LINEARMINUS}) as well. The product of the null dust momenta is%
\begin{align}
\Pi_{+}\Pi_{-} & =\frac{-\left( \Lambda H_{\perp}^{G}\right) ^{2}+\left(
H_{r}^{G}\right) ^{2}}{4\Phi_{+}^{\prime2}\Phi_{-}^{\prime2}}  \notag \\
& =\frac{-\left( \Lambda H_{\perp}^{BH}\right) ^{2}+\left( H_{r}^{BH}\right)
^{2}}{4\Phi_{+}^{\prime2}\Phi_{-}^{\prime2}}\;\;,  \label{PIBYPI}
\end{align}
where we have employed the constraints (\ref{LINEARPLUS}) and (\ref%
{LINEARMINUS}) in the first equality and the constraints (\ref{HBTOTBH}) and
(\ref{HRTOTBH}) in the second equality. Now we can express either $%
H_{\bot}^{BH}$ or $H_{r}^{BH}$ as a square root:%
\begin{align}
H_{\bot}^{BH} & =\sqrt{-\frac{4}{\Lambda^{2}}\Phi_{+}^{\prime}\Phi
_{-}^{\prime}\Pi_{+}\Pi_{-}+g^{rr}H_{r}^{BH}H_{r}^{BH}}\;\;,  \notag \\
H_{r}^{BH} & =\sqrt{\Phi_{+}^{\prime}\Phi_{-}^{\prime}\Pi_{+}\Pi_{-}+\left(
\Lambda H_{\bot}^{BH}\right) ^{2}}\;\;.
\end{align}
These formulae are analogous to the result (\ref{HDbot2}).

The simple linear transformation%
\begin{align}
T & =\frac{1}{\sqrt{2}}\left( \Phi_{+}+\Phi_{-}\right) \,,  \notag \\
\sigma & =\frac{1}{\sqrt{2}}\left( \Phi_{+}-\Phi_{-}\right) \,,  \notag \\
P_{T} & =\frac{1}{\sqrt{2}}\left( \Pi_{+}+\Pi_{-}\right) \,,  \notag \\
P_{\sigma} & =\frac{1}{\sqrt{2}}\left( \Pi_{+}-\Pi_{-}\right)
\end{align}
is a canonical transformation as can be checked by calculating the Poisson
brackets of the new canonical variables $T$, $\sigma$, $P_{T}$, $P_{\sigma}$
. The sum of the constraints (\ref{LINEARPLUS}) and (\ref{LINEARMINUS}),
divided by $\sqrt{2}$, in the new canonical chart becomes%
\begin{equation}
0=P_{T}+\frac{\sigma^{\prime}\Lambda H_{\bot}^{G}+T^{\prime}H_{r}^{G}}{%
T^{\prime2}-\sigma^{\prime2}}\;\;.  \label{PTKENYSZER}
\end{equation}
Similarly the difference of \ (\ref{LINEARPLUS}) and (\ref{LINEARMINUS}),
divided by $\sqrt{2}$, reads%
\begin{equation}
0=P_{\sigma}+\frac{-T^{\prime}\Lambda H_{\bot}^{G}-\sigma^{\prime}H_{r}^{G}}{%
T^{\prime2}-\sigma^{\prime2}}\;\;.  \label{PSZIGKENYSZER}
\end{equation}
The canonical momenta of the cross-flowing null dust are then completely
separated from the rest of the variables in the new constraints (\ref%
{PTKENYSZER}) and (\ref{PSZIGKENYSZER}).

If the null vectors $\,\,l_{\pm~a}\propto\Phi_{\pm,a}$ are both
future-oriented, then $T$ is time-like coordinate. Therefore the
quantization of the constraint (\ref{PTKENYSZER}) will give a functional Schr%
\"{o}dinger equation.

\subsection{Static case}

In this subsection we establish the connection between the canonical
variables ($\Phi_{\pm},\,\Pi_{\pm}$) employed in Ref. \cite{BicakHajicek}
and our canonical coordinates ($Z,\,L$) and momenta ($P_{Z},\,P_{L}$). In
order to do this, first we introduce the ''tortoise-type'' radial coordinate 
$R^{\ast}$ defined as $dR^{\ast}=R\left( L\right) dL$. Next we define null
coordinates $X_{\pm}=Z\pm R^{\ast}$ (see \cite{Gergely}).

As in the static scenario the metric is uniquely given by Eq. (\ref{ds2}),
we would like to identify the energy-momentum tensors (\ref{BHTENZOR}) and (%
\ref{enimp}), which yields $l_{+}^{a}=\kappa^{2}u^{a}/R^{2}W$ and $%
l_{-}^{a}=\kappa^{2}v^{a}/R^{2}W$, with a proportionality constant $%
\kappa^{2}$. According to Ref. \cite{BicakHajicek} the null forms $%
l_{\pm}=\lambda ^{2}\left( \left| \Pi_{\pm}\right| /\sqrt{g}\right)
d\Phi_{\pm}$ ($\lambda^{2}$ a possible second proportionality constant). By
employing Eq. (\ref{uv}), we conclude that%
\begin{equation}
\Pi_{\pm}d\Phi_{\pm}=\mp\sqrt{g}\left( \frac{\kappa}{\lambda}\right) ^{2}%
\frac{dX_{\pm}}{R^{2}}\,,  \label{transf0}
\end{equation}
with $R$ regarded as a function of $X_{\pm}$.

Equivalently, the derived canonical coordinates ($T,\,\sigma$) are related
in a simple way to ($Z,\,L$):

\begin{align}
P_{\sigma}dT+P_{T}d\sigma & =-\sqrt{g}\left( \frac{2\kappa}{\lambda}\right)
^{2}\frac{dZ}{R^{2}}\,,  \notag \\
P_{T}dT+P_{\sigma}d\sigma & =\sqrt{g}\left( \frac{2\kappa}{\lambda}\right)
^{2}\frac{dL}{R}\,,  \label{transf}
\end{align}
with $R$ representing the function (\ref{RL}) of $L$.

The transformations (\ref{transf0}) and (\ref{transf}) establish the
relation between our results and the results derived in Ref. \cite%
{BicakHajicek}.

\section{Concluding Remarks}

The geometrical optics approximation of radiation fields is represented by
null dust. This approximation is a very good one whenever the wavelength of
the radiation is short as compared to the typical local curvature scale of
the space-time. The crossflow of two such radiation streams describes the
interesting situation of a two-component radiation atmosphere of a star. The
assumptions of spherical symmetry and staticity lead to the space-time (\ref%
{ds2}). The two null dust components interact only gravitationally (through
the curvature of the space-time they jointly produce) and formally they are
equivalent to an anisotropic fluid.

A previous canonical treatment of such a system \cite{BicakHajicek}, besides
its many achievements, still suffers from the lack of an internal time (a
difficulty first encountered in the case of one-component null dust). The
existence of such a preferred time function would simplify the quantization
of the gravitational field in question. The absence of the internal time
from the formalism of Ref. \cite{BicakHajicek} is not a major inconvenience
for the analysis presented there, where the one-component null dust limit
(Vaidya space-time) is discussed in detail.

If one does not aim to have this limit in the formalism, the situation is
different. We have shown on the example of the static crossflow of radiation
streams how to construct an internal time for the two-component system. By
suitable canonical transformations we have introduced the time function $Z$
as canonical coordinate and we have constructed the new super-Hamiltonian
and super-momentum constraints, Eqs. (\ref{LINEARZ}), (\ref{LINEARL}), which
have strongly vanishing Poisson brackets. With this, we have turned the
Dirac algebra of the original constraints into an Abelian algebra.

The new constraints contain the momenta conjugate to the crossflowing null
dust variables linearly. This convenient feature can be further exploited in
the process of quantization, which will turn the new super-Hamiltonian
constraint into a functional Schr\"{o}dinger equation. The latter has the
obvious advantage over the Wheeler-deWitt equation obtained by the
quantization of the original super-Hamiltonian constraint, that its space of
solutions is linear. Further properties of the resulting functional Schr\"{o}%
dinger equation are under investigation and we propose to discuss this topic
in detail elsewhere.

\section*{Acknowledgments}

This work was supported by OTKA grants no. T046939 and TS044665. L\'{A}G
wishes to thank the support of the J\'{a}nos Bolyai Scholarship of the
Hungarian Academy of Sciences.

\end{document}